\documentclass[aps,prl,twocolumn,groupedaddress,showpacs]{revtex4-1}
\usepackage{amsmath}
\usepackage{amsfonts,amssymb, graphicx}
\usepackage{wrapfig}
\usepackage[usenames,dvipsnames,svgnames,table]{xcolor}

\begin{document}
\pagestyle{headings}

\title{{Dual Kinetic Balance Approach to the Dirac Equation for Axially Symmetric Systems:
Application to Static and Time-Dependent Fields}}
\author{E.B. Rozenbaum}
\email[]{e.rozenbaum@pcqnt1.phys.spbu.ru}
\affiliation{Department of Physics, St. Petersburg State University, $198504$ St. Petersburg, Russia}
\author{D.A. Glazov}
\affiliation{Department of Physics, St. Petersburg State University, $198504$ St. Petersburg, Russia}
\author{V.M. Shabaev}
\affiliation{Department of Physics, St. Petersburg State University, $198504$ St. Petersburg, Russia}
\author{K.E. Sosnova}
\affiliation{Department of Physics, St. Petersburg State University, $198504$ St. Petersburg, Russia}
\author{D.A. Telnov}
\affiliation{Department of Physics, St. Petersburg State University, $198504$ St. Petersburg, Russia}

\date{\today}

\begin{abstract}
Dual kinetic balance (DKB) technique was previously developed to eliminate spurious states in the finite-basis-set-based solution of the Dirac equation in central fields. In the present paper, it is extended to the Dirac equation for systems with axial symmetry. The efficiency of the method is demonstrated by the calculation of the energy spectra of hydrogenlike ions in presence of static uniform electric or magnetic fields. In addition, the DKB basis set is implemented to solve the time-dependent Dirac equation making use of the split-operator technique. The excitation and ionization probabilities for the hydrogenlike argon and tin ions exposed to laser pulses are evaluated.
\end{abstract}


\maketitle

\section{Introduction}
The finite-basis-set methods are widely used in atomic, molecular and solid state physics.
These methods generally possess high level of the numerical efficiency that includes a fast growth of the accuracy with increasing number of basis functions.
It is known, however, that the straightforward application of this kind of methods to the Dirac equation leads to appearance of so-called spurious states (see Refs. \cite{bra77,dra81,kut84,schl,joh86,joh88,sha04,tup08,lew10,lewarx} and references therein).
As it was shown in Refs. \cite{sha04,tup08}, the spurious states originate from the restriction of the basis set to a finite number of functions. Several methods have been developed to solve this problem in the case of spherical symmetry. Among them is the well-known kinetic balance method \cite{lee82,gra82,dya84,sta84,qui87} that implies the construction of the lower-component basis functions by applying the non-relativistic limit of the radial Dirac operator to the upper component basis functions. In Refs. \citep{ish87,ish97}, the basis was composed of the Gaussian spinors satisfying the boundary conditions for the case of the finite nucleus. The Gaussian spinors obey the kinetic-balance condition (except for the non-relativistic limit). Advantage of these approaches is a high accuracy in calculating the bound-state energies. However, since the kinetic balance method violates the symmetry in treatment of the positive- and negative-energy states, application of this technique can be rather problematic if the contribution of the negative-energy continuum is significant. In particular, it takes place in calculations of the QED effects (e.g. for precise calculations of the $g$-factor). An effective and easily implementable method to get rid of the spurious states keeping the symmetry between the electron and the positron states was proposed in Ref. \cite{sha04}. This method is called the dual kinetic balance (DKB) approach. The efficiency of the DKB method was proved by the calculations of various relativistic and QED effects in atomic systems \cite{sha05, vol09, bel09, gla10, yer11, sun11, che11, zat12, mcc12, OngBF}. In the present work, the DKB method is generalized to the Dirac equation for systems with axial symmetry.

At first, the finite-basis-set method is developed for stationary Dirac equation with axially symmetric potential. The basis set is constructed from the one-component basis functions of radial and angular variables and transformed to the DKB form.
To test the procedure, the spectra of hydrogenlike ions are obtained with making no use of spherical symmetry of these ions. It is shown that the DKB approach allows one to get rid of the spurious states while retaining the proper energy spectrum.
To demonstrate the efficiency of the method, the spectra of hydrogenlike ions in strong uniform electric and magnetic fields  are calculated. The DKB approach does eliminate the spurious solutions in these cases as well. The results for the Zeeman and Stark shifts of the levels with $n = 1$ for $Z=1,18,50,$ and $92$ are compared with those of the independent relativistic calculations based on the perturbation theory. 

The developed method has a wide range of possible applications. In particular, it can be used for calculations of the Zeeman effect, including linear ($g$ factor) and non-linear contributions in magnetic fields. The latter appears to be important for middle-$Z$ boronlike ions at the present level of experimental accuracy \cite{lindenfels:13:pra}. In this respect, the DKB method represents a competitive alternative to the traditionally employed perturbation theory.

The time-dependent problems have been drawing much more attention during last years due to the rapid development of the laser technologies. There are several state-of-the-art laser facilities operating nowadays (see, e.g., \cite{vogel12}) that provide extremely high intensities or frequencies of the radiation. Thus the processes involving the strong-field ionization and excitation are of a great interest \cite{sorok07, rich09, Lindroth, Pindzola1, Pindzola2, Saenz}. Highly charged ions are among the most interesting objects that can be experimentally studied with these lasers. Theoretical treatment of highly charged ions exposed to strong laser fields requires the fully relativistic consideration. Thus the time-dependent Dirac equation is to be solved. Within the mostly relevant dipole approximation and for the linearly polarized laser fields, these problems possess the axial symmetry and the solution can be based on the approach developed in the present paper for the stationary Dirac equation. The particular scheme of the solution of the time-dependent Dirac equation is based on the split-operator technique and requires some transformations of the matrices. The developed technique is applied to calculate the excitation probabilities for hydrogenlike argon ion and the ionization probabilities for hydrogenlike tin ion exposed to various intense laser pulses. Some of the obtained results for the excitation probabilities are compared with those of the independent calculation (based on the first-order non-stationary perturbation theory). The results for the ionization probability dependence on the laser wavelength are compared with the corresponding data from Ref. \cite{Saenz}.

Throughout the paper we assume $\hbar = 1$.

\section{Stationary Dirac equation}
We consider the stationary Dirac equation
  \begin{equation} \label{Dir1}
    H\Psi(\mathbf{r}) = E\Psi(\mathbf{r}),
  \end{equation}
where
\vspace{-10pt}
  \begin{equation}
    H = c(\boldsymbol{{\alpha}}\cdot\mathbf{{p}}) + mc^2{\beta} + V \,,
  \end{equation}
\vspace{-17pt}
  \begin{equation}
  \label{potential}
    V = V_\mathrm{nucl}(r) + V_\mathrm{ext}(r, \theta) \,.
  \end{equation}
Here $V_\mathrm{nucl}(r)$ is the nuclear potential. The stationary external field potential $V_\mathrm{ext}(r, \theta)$, being dependent on the radial and angular variables ($r$ and $\theta$), takes different forms for electric and magnetic fields. In the case of external electric field, it is given in the length gauge by:
  \begin{equation}
   \label{eq:VE}
    V_\mathrm{ext} = -(\boldsymbol{\mathcal{E}}\cdot\mathbf{d}) = -e\left(\boldsymbol{\mathcal{E}}\cdot\mathbf{r}\right) \,,
  \end{equation}
where $e$ is the electron charge and $\mathbf{d}$ is the dipole moment operator: $\mathbf{d} = e\mathbf{r}$.
For external magnetic field, we have
  \begin{equation}
    V_\mathrm{ext} = -e\left(\boldsymbol{\alpha}\cdot\mathbf{A}\right) \,,
  \end{equation}
which can be rewritten in the particular gauge, $\mathbf{A}=\dfrac{1}{2}\left[\boldsymbol{\mathcal{H}}\times\mathbf{r}\right]$, as:
  \begin{equation}
   \label{eq:VH}
    V_\mathrm{ext} = -\dfrac{e}{2}\left(\boldsymbol{\left[\mathbf{r}\times\boldsymbol{\alpha}\right]\cdot\mathcal{H}}\right) \,.
  \end{equation}
Here and below the external field is assumed to be directed along the $z$ axis: $\boldsymbol{\mathcal{E}},\boldsymbol{\mathcal{H}}\upuparrows \mathbf{e}_z$; $r$, $\theta$, and $\varphi$ are the corresponding spherical coordinates. In the case of an axially symmetric field $V(r, \theta)$, the total angular momentum $J$ is not conserved. At the same time, the $z$-projection of the total angular momentum $m_{_J}$ is conserved because the corresponding operator $J_z$ commutes with the Hamiltonian:
\begin{equation}
  \left[J_z, H\right] = 0 \,.
\end{equation}
Consequently, {$H$ and $J_z$} have a common set of eigenfunctions with explicit dependence on the azimuthal angle $\varphi$ and thus the Dirac four-component wave function (bispinor) can be represented in the spherical coordinates as follows:
  \begin{equation} \label{PsiFunc}
    \Psi(\textbf{r}) = \dfrac{1}{r}
    \left(
      \begin{array}{rcl}
        G_1(r, \theta)\hspace{1pt}\mathrm{e}^{i(m_{_J}-\frac{1}{2})\varphi} \vspace{1pt} \\
        G_2(r, \theta)\hspace{1pt}\mathrm{e}^{i(m_{_J}+\frac{1}{2})\varphi} \vspace{1pt} \\
        \dot{\imath}F_1(r, \theta)\hspace{1pt}\mathrm{e}^{i(m_{_J}-\frac{1}{2})\varphi} \vspace{1pt} \\
        \dot{\imath}F_2(r, \theta)\hspace{1pt}\mathrm{e}^{i(m_{_J}+\frac{1}{2})\varphi} \\
      \end{array}
    \right) .
  \end{equation}
Substitution of the form (\ref{PsiFunc}) into the Dirac equation (\ref{Dir1}) yields the equation
  \begin{equation} \label{DirEq}
    H_{m_{_J}}\Phi = E\Phi
  \end{equation}
for the function
  \begin{equation}
    \Phi(r, \theta) =
      \left(
        \begin{array}{rcl}
          G_1(r, \theta)\\
          G_2(r, \theta)\\
          F_1(r, \theta)\\
          F_2(r, \theta)\\
        \end{array}
      \right) \,.
  \end{equation}

In the case of external electric field, the operator $H_{m_{_J}}$ takes the following form:
  \begin{equation} \label{Hamiltonian1}
    H_{m_{_J}} =
      \begin{pmatrix}
        mc^2+V & cD_{m_{_J}} \\
        -cD_{m_{_J}} & -mc^2+V
      \end{pmatrix}\,,
  \end{equation}
where $V$ is given by Eqs.~(\ref{potential}) and (\ref{eq:VE}),
  \begin{eqnarray} \label{OperD}
    D_{m_{_J}} &=& \left(\sigma_z\cos\theta + \sigma_x\sin\theta\right)\left(\dfrac{\partial}{\partial{r}}-\dfrac{1}{r}\right) \nonumber\\
    && + \dfrac{1}{r}\left(\sigma_x\cos\theta - \sigma_z\sin\theta\right)\dfrac{\partial}{\partial\theta} \nonumber\\
    && + \dfrac{1}{r\sin\theta}\left(im_{_J}\sigma_y + \dfrac{1}{2}\sigma_x\right)\,,
  \end{eqnarray}
$\sigma_x, \sigma_y,$ and $\sigma_z$ are the Pauli matrices.

In the case of external magnetic field, the Hamiltonian $H_{m_{_J}}$ has the following form:
  \begin{equation} \label{Hamiltonian2}
    H_{m_{_J}} =
      \begin{pmatrix}
        mc^2 + V_\mathrm{nucl} & c\left(D_{m_{_J}} + \tilde{D}\right) \\
        -c\left(D_{m_{_J}} + \tilde{D}\right) & -mc^2 + V_\mathrm{nucl}
      \end{pmatrix} \,,
  \end{equation}
where
  \begin{eqnarray} \label{OperMagnD}
    \tilde{D} &=& -\dfrac{e}{2c}\mathcal{H}r\sin\theta\,\dot{\imath}\sigma_y\,.
  \end{eqnarray}

We note that $D_{m_{_J}}$ and $\tilde{D}$ are anti-Hermitian operators:
  \begin{equation}
    D_{m_{_J}}^\dagger = -D_{m_{_J}}\,,
  \end{equation}
\vspace{-17pt}
  \begin{equation}
    \tilde{D}^\dagger = -\tilde{D}\,.
  \end{equation}

The scalar product in the space of the functions $\Phi$ is defined by
  \begin{eqnarray}
    \langle \Phi^a|\Phi^b\rangle &=& \int\limits_0^\infty{dr}\int\limits_0^\pi{d\theta\sin\theta}
    (  G_1^aG_1^b + G_2^aG_2^b \nonumber\\
    && + F_1^aF_1^b + F_2^aF_2^b ) \,.
  \end{eqnarray}
Setting the boundary conditions,
  \begin{equation}
    \left.\Phi(r, \theta)\right|_{r = 0} = \lim\limits_{r \rightarrow \infty}\Phi(r, \theta) = 0 \,,
  \end{equation}
leads Eq. (\ref{DirEq}) to be equivalent to the variational principle $\delta{\mathbb{S}}=0$ for the functional
  \begin{equation} \label{functional}
    \mathbb{S} =  \left<\Phi|H_{m_{_J}}|\Phi\right> - E\left<\Phi|\Phi\right> \,,
  \end{equation}
where the undefined Lagrange factor $E$ has the physical meaning of energy.  \\

Implementation of any kind of methods based on finite basis sets starts with an approximate representation of unknown function $\Phi$ as a finite linear combination of the basis functions.
Let $N$ be the number of the four-component basis functions depending on the radial and angular variables ($r$ and $\theta$). We introduce a set of functions $\left.\{W_i(r, \theta)\}\right|_{i=1}^{N}$, where $r \in [0, r_\mathrm{max}]$ and $\theta \in [0, \pi]$.
Then the function $\Phi$ can be expanded as follows:
  \begin{equation} \label{tv_expansion}
    \Phi(r, \theta) \cong \sum\limits_{i=1}^NC_iW_i(r, \theta) \,,
  \end{equation}
where $C_i$ are the expansion coefficients.

By substitution of the expansion (\ref{tv_expansion}) into the variational principle $\delta{\mathbb{S}} = 0$, the latter can be represented as a set of algebraic equations for the coefficients $C_i$:
  \begin{eqnarray} \label{diffEigVal}
    \dfrac{d\mathbb{S}}{dC_i} = 0 \,.
  \end{eqnarray}
This system leads to the following generalized eigenvalue problem:
  \begin{equation} \label{GenEigValProb}
    H_{ij}C_j=
    ES_{ij}C_j \,,
  \end{equation}
where the summation over the repeated indices is implied,
  \begin{eqnarray} \label{H0Matrix}
    H_{ij}
    = \int\limits_0^{\infty}dr\int\limits_0^{\pi} {d}\theta\sin\theta
      \left[W_i(r, \theta)\right]^\dagger
      H_{m_{_J}}W_j(r, \theta) \,,
  \nonumber\\
  \end{eqnarray}
  \begin{eqnarray} \label{S0Matrix}
    S_{ij}
    = \int\limits_0^{\infty}dr\int\limits_0^{\pi}{d}\theta \sin\theta
      \left[W_i(r, \theta)\right]^\dagger
      W_j(r, \theta) \,,
  \nonumber\\
  \end{eqnarray}
and the Hamiltonian $H_{m_{_J}}$ is defined by Eq.~(\ref{Hamiltonian1}) or by Eq.~(\ref{Hamiltonian2}).

Consider the construction of the basis set. Let $N_r$ and $N_\theta$ be the numbers of the one-component basis functions depending on the $r$ and $\theta$ variables, respectively. We denote these sets of functions as $\{B_{i_r}(r)\}_{i_r=1}^{N_r}$ and $\{Q_{i_\theta}(\theta)\}_{i_\theta=1}^{N_\theta}$. The indices $i_r = 1, \ldots, N_r$,\; \vspace{3pt} $i_\theta = 1, \ldots, N_\theta$,\; and $u = 1, \ldots, 4$ compose a single index $i=1,\ldots, N$ introduced before ($N=4N_rN_\theta$) as follows:
\begin{eqnarray}
i = (u-1)N_rN_\theta + (i_\theta-1)N_\theta + i_r \,.
\end{eqnarray}
Using these one-component single-variable function sets, we can construct the set of four-component functions $W_i(r, \theta) = W^{(u)}_{{i_r}{i_\theta}}(r, \theta)$ of two variables. Then the expansion (\ref{tv_expansion}) will take the form:
  \begin{equation} \label{tv_expansion_2}
    \Phi(r, \theta) \cong \sum\limits_{u=1}^4\sum\limits_{i_r=1}^{N_r}\sum\limits_{i_\theta=1}^{N_\theta}C^u_{{i_r}{i_\theta}}W^{(u)}_{{i_r}{i_\theta}}(r, \theta) \,,
  \end{equation}
and indices $i$ and $j$ in Eqs. (\ref{diffEigVal}) - (\ref{S0Matrix}) should be replaced with $\{i_r, i_\theta, u\}$ and $\{j_r, j_\theta, v\}$, respectively.

A straightforward way to construct the four-component basis functions depending on two variables ($r$ and $\theta$)
is:
  \begin{equation} \label{basis_stand}
    W^{(u)}_{{i_r}{i_\theta}}(r, \theta) = B_{i_r}(r)Q_{i_\theta}(\theta)\;\mathrm{e}_u, 
  \end{equation}
where
  \begin{equation}
  \hspace{-1pt}
    \mathrm{e}_1 = \left(
        \begin{array}{rcl}
          1\\
          0\\
          0\\
          0
        \end{array}
      \right) \hspace{-3pt},\;
    \mathrm{e}_2 = \left(
        \begin{array}{rcl}
          0\\
          1\\
          0\\
          0
        \end{array}
      \right)\hspace{-3pt},\;
    \mathrm{e}_3 = \left(
        \begin{array}{rcl}
          0\\
          0\\
          1\\
          0
        \end{array}
      \right)\hspace{-3pt},\;
    \mathrm{e}_4 = \left(
        \begin{array}{rcl}
          0\\
          0\\
          0\\
          1
        \end{array}
      \right)\hspace{-3pt}.
      \hspace{-7pt}
  \end{equation}
Our calculations with various finite-basis-set techniques, including
the B-splines-based spectral approach \cite{schl,joh86,joh88} and the generalized pseudo-spectral method \cite{GPS}, show that the basis  (\ref{basis_stand}) leads to the appearance of the spurious states.

Following the idea of the DKB method, we should impose specific relations between the upper and lower components of the Dirac bispinor. These relations are derived from the non-relativistic limit of the Dirac equation and give the following basis functions in the case of axial symmetry:
  \begin{equation}
  \label{eq:DKB}
    W^{(u)}_{{i_r}{i_\theta}}(r, \theta) = {\Lambda}B_{i_r}(r)Q_{i_\theta}(\theta)\;\mathrm{e}_u, \;\;\; u = 1, \ldots, 4 \,,
  \end{equation}
where
  \begin{equation}
    \Lambda =
    \begin{pmatrix}
      1 & -\dfrac{1}{2mc}D_{m_{_J}} \vspace{10pt}\\
      -\dfrac{1}{2mc}D_{m_{_J}} & 1
    \end{pmatrix}.
  \end{equation}

It should be noted that, as in the case of central fields \cite{sha04}, the DKB approach for axially symmetric systems can be used for the extended charge nucleus only. The point-like nucleus case can be accessed by the extrapolation of the extended-nucleus results to vanishing nuclear size.

The discussion above is given for arbitrary basis sets $\{B_{i_r}(r)\}_{i_r=1}^{N_r}$ and $\{Q_{i_\theta}(\theta)\}_{i_\theta=1}^{N_\theta}$. In the present work, the particular choice of the one-component basis functions is made as follows. The $B$-splines of some order $k$ form the set of the one-component $r$-dependent basis functions{, $\{B_{i_r}(r)\}_{{i_r}=1}^{N_r}$}. The Legendre polynomials {$\left\{P_l\left(\dfrac{2}{\pi}\theta-1\right)\right\}_{{l}=0}^{N_\theta-1}$} of degrees {$l = 0, \ldots N_\theta-1$} form the set of the one-component $\theta$-dependent basis functions, so that in the previous notations $Q_{i_\theta}(\theta) \equiv P_{i_\theta-1}\left(\dfrac{2}{\pi}\theta-1\right)$.

To demonstrate the absence of the spurious states in calculations based on the DKB method, in Table \ref{SpectrumTin} we present the energy spectrum of hydrogenlike tin ion ($Z=50$), evaluated for the extended nucleus case (the model of the uniformly charged sphere is employed) with the plain basis set~(\ref{basis_stand}) and with the DKB basis set~(\ref{eq:DKB}).
The calculations are performed for the projection of the total angular momentum $m_{_J} = -1/2$ making no use of the spherical symmetry of the ion. For comparison, the exact values (within the indicated digits, see Table \ref{SpectrumTin}) obtained by the numerical solution of the radial Dirac equation using the finite difference method are presented as well. It can be noticed that the DKB method eliminates the spurious states, maintaining the same level of accuracy for the energies.

In order to prove the applicability of the present method to the case of axially symmetric fields, we have calculated the energy spectra of hydrogenlike ions in presence of static uniform electric or magnetic fields. Tables \ref{ZeemanHydB=0.1} and \ref{ZeemanHydB=123} present Zeeman shifted levels in hydrogen atom exposed to different magnetic fields. They are compared with the corresponding results from Refs. \cite{goldman, rutpos, naknak} and with the results of our independent perturbation theory (PT) calculations. Table \ref{Zeeman} presents the Zeeman shifts of the energy levels for hydrogenlike argon, tin, and uranium ions in different magnetic fields. For comparison, we have also evaluated these shifts within the perturbation theory, where the zero-order approximation corresponds to the Dirac equation with the nuclear potential, and the interaction with the external magnetic field, given by Eq. (\ref{eq:VH}), is treated perturbatively. It has been implemented numerically as an iterative procedure, where the energies and the wave functions of the $n$-th order are computed from the energies and the wave functions of the $(n-1)$-th order. The summations over the spectrum have been performed with the help of the DKB finite-basis-set method for spherically symmetric case \cite{sha04}. Data in Table \ref{ZeemanHydB=0.1} show that our method reproduces as many orders of the perturbation theory as needed up to its own numerical accuracy.

Tables \ref{StarkHydTin} and \ref{StarkTin2} display the Stark shifts for hydrogen atom and hydrogenlike argon ion in uniform electric fields. We have calculated the Stark effect within the perturbation theory as well. The results for the expansion coefficients are in agreement with those obtained in Refs.~\cite{Mendelson,OvsPal}. Furthermore, our values are obtained within the relativistic treatment and are valid to all orders in $\alpha Z$ in any order of the field strength. Tables \ref{StarkHydTin} and \ref{StarkTin2} show that the present DKB approach fully reproduces the perturbation theory results. In this case, however, one should keep in mind that, strictly speaking, there are no discrete energy levels for atom in uniform electric field. Instead, we have the quasi-stationary states. It happens due to the tunneling effect for initially localized electron state.

The basis set of $78$ radial B-splines of order $k=9$ and $17$ Legendre polynomials (of orders from $0$ to $16$) is enough to obtain all the results presented in Tables \ref{SpectrumTin} $-$ \ref{StarkTin2}.

\begin{table}[htbp]
\caption{{Energy spectrum (in r.u.) of H-like tin ion\\
($Z=50$, $R_\mathrm{nucl}=4.655$ fm).}}
\begin{ruledtabular}
\begin{tabular}{cccc}
n & DKB off & DKB on & Exact values \\ \hline
\vspace{-8pt} \\
1 & 0.93106324090 & 0.93106324090 & 0.93106324086 \\
 & 0.97072224116 &  &  \\ \hline
 \vspace{-8pt} \\
2 & 0.98261372423 & 0.98261372423 & 0.98261372423 \\
 & 0.98261424969 & 0.98261424969 & 0.98261424969 \\
 & 0.98321813638 & 0.98321813638 & 0.98321813638 \\
 & 0.98659670113 &  &  \\ \hline
 \vspace{-8pt} \\
3 & 0.99234087351 & 0.99234087351 & 0.99234087351 \\
 & 0.99234102938 & 0.99234102938 & 0.99234102937 \\
 & 0.99252042806 & 0.99252042806 & 0.99252042806 \\
 & 0.99252042806 & 0.99252042806 & 0.99252042806 \\
 & 0.99257642386 & 0.99257642386 & 0.99257642386 \\
 & 0.99302522647 &  &  \\
\end{tabular}
\end{ruledtabular}
\label{SpectrumTin}
\end{table}

\begin{table}[htbp]
\caption{Binding energy (in~a.u.) of the ground state ($m_{_J} = -1/2$) of hydrogen atom in uniform magnetic field $\mathcal{H}=0.1$ a.u. ($\approx 2.35 \cdot 10^4$ T). For comparison, the value obtained in Refs.~\cite{goldman,rutpos} (they both coincide to all the presented digits) is given. The complete perturbation-theory result $\Delta{E}_\mathrm{PT}$ and the individual contributions are listed as well. The terms missing in the breakdown (the odd orders $>3$ and the even orders $>12$) are zero to all the presented digits.}
\begin{ruledtabular}
\begin{tabular}{ccrc}
This work & {Refs. \cite{goldman, rutpos}} & PT order & {PT} \\ \hline
\vspace{-8pt} \\
$-0.547532408$ & $-0.547532408$ & (up to 12) & $-0.547532410$ \\
 &  & (1) & $-0.049999114$ \\
 &  & (2) & \hspace{4pt} $0.002499823$ \\
 &  & (3) & \hspace{4pt} $0.000000018$ \\
 &  & (4) & $-0.000027603$ \\
 &  & (6) & \hspace{4pt} $0.000001211$ \\
 &  & (8) & $-0.000000098$ \\
 &  & (10) & \hspace{4pt} $0.000000012$ \\
 &  & (12) & $-0.000000002$ \\
\end{tabular}
\end{ruledtabular}
\label{ZeemanHydB=0.1}
\end{table}

\begin{table}[htbp]
\caption{Binding energies (in~a.u.) of the ground ($m_{_J}=-1/2$) state and the lowest $m_{_J}=-3/2$ state of hydrogen atom in uniform magnetic field $\mathcal{H}$. }
\begin{ruledtabular}
\begin{tabular}{ccll}
$\mathcal{H}$, a.u. & $m_{_J}$ &This work & Refs. \cite{goldman, rutpos, naknak} \\ \hline
\vspace{-8pt} \\
$1$ & $-1/2$ & $-0.8311725$ & $-0.8311732^{abc}$ \\
&$-3/2$ & $-0.456592$ & $-0.456597^{ac}$ \\ \hline
\vspace{-8pt} \\
$2$ & $-1/2$ & $-1.022216$ & $-1.022218^{ab}$ \\ \hline
\vspace{-8pt} \\
$3$ & $-1/2$ & $-1.164528$ & $-1.164537^{ab}$ \vspace{-2pt}
\end{tabular}
\end{ruledtabular}
\label{ZeemanHydB=123}
Taken from: $^a$ Ref. \cite{goldman}; $^b$ Ref. \cite{rutpos}; $^c$ Ref. \cite{naknak}.
\end{table}

\begin{table}[htbp]
\caption{Zeeman shifts (in~a.u.) of the ground states of H-like ions. $\Delta{E}_\mathrm{PT}$ are the perturbation-theory values. The contributions of the orders $>4$ are zero to all the presented digits.}
\begin{ruledtabular}
\begin{tabular}{ccrc}
$Z = 18$ & $R_\mathrm{nucl}=3.427$ fm & $\mathcal{H} = 6\cdot10^5$ T &  \\ \hline
\vspace{-8pt} \\
$m_{_J}$ & $\Delta{E}_\mathrm{DKB}$ & {PT order} & $\Delta{E}_\mathrm{PT}$ \\ \hline
\vspace{-8pt} \\
$+1/2$ & \hspace{4pt} $1.27385376$ & (up to 4) & \hspace{4pt} $1.27385371$ \\
$-1/2$ & $-1.26402973$ & (up to 4) & $-1.26402967$ \\
$\pm1/2$ &  & (1) & $\pm1.26894260$ \\
$\pm1/2$ &  & (2) & \hspace{7.6pt}$0.00491236$ \\
$\pm1/2$ &  & (3) & $\mp0.00000091$ \\
$\pm1/2$ &  & (4) & $-0.00000034$ \\ 
\hline\hline \vspace{-8pt} \\ 
$Z=50$ & $R_\mathrm{nucl}=4.655$ fm & $\mathcal{H}=6\cdot10^6$ T &  \\ \hline
\vspace{-8pt} \\
$m_{_J}$ & $\Delta{E}_\mathrm{DKB}$ & PT order & $\Delta{E}_\mathrm{PT}$ \\ \hline
\vspace{-8pt} \\
$+1/2$ & \hspace{4pt} $12.2303556$ & (up to 4) & \hspace{4pt} $12.2303551$ \\
$-1/2$ & $-12.1227200$ & (up to 4) & $-12.1227195$ \\
$\pm1/2$ &  & (1) & $\pm12.1766415$ \\
$\pm1/2$ &  & (2) & \hspace{9pt} $0.0538243$ \\
$\pm1/2$ &  & (3) & \hspace{1.8pt} $\mp0.0001042$ \\
$\pm1/2$ &  & (4) & \hspace{1.8pt} $-0.0000065$ \\ 
\hline\hline \vspace{-8pt} \\ 
$Z=92$ & $R_\mathrm{nucl}=5.8569$ fm & $\mathcal{H}=6\cdot10^7$ T &  \\ \hline
\vspace{-8pt} \\
$m_{_J}$ & $\Delta{E}_\mathrm{DKB}$ & PT order & $\Delta{E}_\mathrm{PT}$ \\ \hline
\vspace{-8pt} \\
$+1/2$ & \hspace{4pt} $106.5190$ & (up to 4) & \hspace{4pt} $106.5188$ \\
$-1/2$ & $-104.8140$ & (up to 4) & $-104.8138$ \\
$\pm1/2$ &  & (1) & $\pm105.6865$ \\
$\pm1/2$ &  & (2) & \hspace{13.6pt} $0.8534$ \\
$\pm1/2$ &  & (3) & \hspace{6pt} $\mp0.0202$ \\
$\pm1/2$ &  & (4) & \hspace{6pt} $-0.0009$ \\
\end{tabular}
\end{ruledtabular}
\label{Zeeman}
\end{table}

\begin{table}[htbp]
\caption{Stark shifts (in~a.u.) of the ground state of hydrogen atom and H-like argon ion. $\Delta{E}_\mathrm{PT}$ and $\Delta{E}^\mathrm{nr}_\mathrm{PT}$ are the relativistic and non-relativistic perturbation-theory values. The contributions beyond the shown ones are zero to all the presented digits.}
\begin{center}
\begin{ruledtabular}
\begin{tabular}{ccrcc}
&& $Z=1$ && \\ \hline
\vspace{-8pt} \\ 
$\mathcal{E}$, V/m & $\Delta{E}_\mathrm{DKB}\cdot{10^6}$ & PT order & {$\Delta{E}_\mathrm{PT}\cdot{10^6}$} & {$\Delta{E}^\mathrm{nr}_\mathrm{PT}\cdot{10^6}$} \\ \hline
\vspace{-8pt} \\ 
$5\cdot10^8$ &  $-2.1271$ & (2) & $-2.1272$ &$-2.1273$\\
\hline
\vspace{-8pt} \\ 
$1\cdot10^9$ & $-8.50944$ & (up to 4) & $-8.50942$ & $-8.50989$\\
 & &(2) & $-8.50863$ &$-8.50910$\\
 & & (4) & $-0.00079$ & $-0.00079$ \\ \hline
\vspace{-8pt} \\ 
$2\cdot10^9$ & $-34.0472$ & (up to 4) & $-34.0472$ &$-34.0491$\\
 & &(2) & $-34.0345$ &$-34.0364$\\
 & &(4) & \hspace{2pt} $-0.0127$ &\hspace{2pt} $-0.0127$\\
 \hline\hline \vspace{-8pt} \\ 
&& $Z=18$ & $R_\mathrm{nucl}=3.427$ fm & \\ \hline
\vspace{-8pt} \\ 
$\mathcal{E}$, V/m & $\Delta{E}_\mathrm{DKB}\cdot10^6$ & PT order & {$\Delta{E}_\mathrm{PT}\cdot10^6$} & {$\Delta{E}^\mathrm{nr}_\mathrm{PT}\cdot10^6$} \\ \hline
\vspace{-8pt} \\ 
$5\cdot10^{11}$ & $-19.9026$ & (2) & $-19.9027$ & $-20.2644$ \\
$1\cdot10^{12}$ & $-79.6105$ & (2) & $-79.6108$ & $-81.0578$ \\ 
$2\cdot10^{12}$ & \hspace{-7pt} $-318.4454$ & (2) & \hspace{-7pt} $-318.4460$ & \hspace{-7.4pt} $-324.2338$ \\
\end{tabular}
\end{ruledtabular}
\end{center}
\label{StarkHydTin}
\end{table}

\begin{table}[htbp]
\caption{Stark shifts (in~a.u.) of the $n=2$ energy levels of H-like argon ion ($Z=18$, $R_\mathrm{nucl}=3.427$~fm). $\Delta{E}_\mathrm{PTD}$ are calculated according to the approximate formulas from Ref.~\cite{zap78} derived within perturbation theory for degenerate levels.}
\begin{center}
\begin{ruledtabular}
\begin{tabular}{cccc}
{$\mathcal{E}$, V/m}& Level & {$\Delta{E}_\mathrm{DKB}$} & {$\Delta{E}_\mathrm{PTD}$} \\ \hline
 & $2s$ & \hspace{4pt} $0.0288$ & \hspace{4pt} $0.0292$ \\
{$2\cdot10^{11}$}& $2p_{1/2}$ & $-0.0441$ & $-0.0444$ \\
& $2p_{3/2}$ & \hspace{4pt} $0.0150$ & \hspace{4pt} $0.0152$ \\ \hline
 & $2s$ & \hspace{4pt} $0.0467$ & \hspace{4pt} $0.0477$ \\
{$5\cdot10^{11}$} & $2p_{1/2}$ & $-0.1267$ & $-0.1268$ \\
&$2p_{3/2}$ & \hspace{4pt} $0.0778$ & \hspace{4pt} $0.0791$ \\
\end{tabular}
\end{ruledtabular}
\end{center}
\label{StarkTin2}
\end{table}

\section{Time-dependent Dirac equation} \label{tdde-solution}
We consider the time-dependent Dirac equation:
      \begin{equation} \label{NonStat}
        \dot{\imath}\dfrac{\partial}{\partial t}\Psi(\mathbf{r}, t) = H(t)\Psi(\mathbf{r}, t) \,,
      \end{equation}
      where
      \vspace{-5pt}
       \begin{equation}
        H(t) = H_0 + \mathcal{V}(t) \,,
      \end{equation}
       \begin{equation} \label{OldH0}
        H_0 = c(\boldsymbol{{\alpha}}\cdot\mathbf{{p}}) + mc^2{\beta} + V_\mathrm{nucl}(\mathbf{r})\,,
      \end{equation}
and $\mathcal{V}(\mathbf{r}, t)$ describes the interaction with an external time-dependent field.
In the following, we restrict our consideration to the time-dependent electric field within the dipole approximation:
\begin{equation}
\label{Vrt}
  \mathcal{V}(\mathbf{r}, t) = -\left(\boldsymbol{\mathcal{F}}(t)\cdot\mathbf{d}\right) \,,
\end{equation}
where $\boldsymbol{\mathcal{F}}$ is the strength of the external electric field and $\mathbf{d}$ is the operator of the dipole moment: $\mathbf{d} = e\mathbf{r}$.
We assume $\boldsymbol{\mathcal{F}}$ to be linearly polarized along the $z$ axis:
\begin{equation} \label{extstrength}
  \boldsymbol{\mathcal{F}}(t) = \mathcal{F}(t)\mathbf{e}_z \,.
\end{equation}

Let ${\Delta}t$ be a small time step. Given the initial wave function $\Psi(\mathbf{r}, 0)$,
 the approximate solution of the time-dependent Dirac equation can be found by iterations:
      \begin{equation}
        \Psi(\mathbf{r}, t+{\Delta}t) \approx \exp\left(-\dot{\imath}{\Delta}tH[\mathbf{r}, t + {\Delta}t/2]\right)\Psi(\mathbf{r}, t) \,.
      \end{equation}
For the function $\Phi$ defined in the previous section, this~equation can be written as
      \begin{eqnarray}
        \Phi(r, \theta; t+{\Delta}t) \approx
        \exp\left(-\dot{\imath}{\Delta}tH_{m_{_J}}[r, \theta; t + {\Delta}t/2]\right)\Phi(r, \theta; t) \,.
\nonumber\\
      \end{eqnarray}
The direct application of these equations within the finite-basis-set approach
would be extremely time consuming. For this reason, one needs to use special
methods to reduce the efforts. We use the split-operator technique \cite{Split-oper}.
The implementation of this technique in the framework of the finite-basis-set method described above requires, however, some modifications that are presented below.

The split-operator method consists in the propagator factorization, e.g., as follows:
    \begin{eqnarray}
      \Psi(\mathbf{r}, t+{\Delta}t) &\approx& \exp\left[-\dot{\imath}\frac{{\Delta}t}{2}H_0\right]
        \exp\left(-\dot{\imath}{\Delta}t\,\mathcal{V}\left[\mathbf{r}, t + \frac{{\Delta}t}{2}\right]\right)
\nonumber\\
      &\times& \exp\left[-\dot{\imath}\frac{{\Delta}t}{2}H_0\right]\Psi(\mathbf{r}, t).
    \end{eqnarray}
The exponential of the unperturbed Hamiltonian $H_0$ is time-independent, and thus can be calculated only once by the spectral expansion:
\begin{equation}\label{exph0}
  \exp\left[-\dot{\imath}\frac{\Delta{t}}{2}H_0\right] =
\sum\limits_k\exp\left(-\dot{\imath}\frac{\Delta{t}}{2}E_k\right)\left|\Psi_k\right>\left<\Psi_k\right| \,,
\end{equation}
where
\begin{equation} \label{OperDir}
 H_0\Psi_k = E_k\Psi_k \,.
\end{equation}
In order to calculate the spectral expansion (\ref{exph0}), we introduce the matrix and eigenvectors:
\begin{equation} \label{orthogonalization}
  H_{0}^L = S^{-1/2}H_{0}S^{-1/2} \,,\;\;
 \vec{C}^L  = S^{1/2}\vec{C} \,,
\end{equation}
so that, instead of the generalized eigenvalue problem (\ref{GenEigValProb}), we get the ordinary one:
\begin{equation} \label{LSimpGenEigValProb}
  H_{0}^L\vec{C}^L = E\vec{C}^L \,.
\end{equation}
In order to get the highest possible efficiency, the time-dependent part $\mathcal{V}\left[\mathbf{r}, t + \frac{{\Delta}t}{2}\right]$ should be represented by a diagonal matrix. According to Eqs.~(\ref{Vrt}) and (\ref{extstrength}), we can represent the matrix $\mathcal{V}$ as
\begin{equation}
  \mathcal{V}(t) = \mathcal{F}(t)\cdot{V}\,,
\end{equation}
where the matrix elements of $V$ are given by
  \begin{eqnarray}
    V^{uv}_{i_ri_{\theta}j_r{j}_\theta}
   & =& \int\limits_0^{\infty}dr\int\limits_0^{\pi}{d}\theta\sin\theta
    \left[W^{(u)}_{i_ri_\theta}(r, \theta)\right]^\dagger \nonumber\\
&&\times\left(r\cos\theta\right)W^{(v)}_{j_rj_\theta}(r, \theta) \,.
  \end{eqnarray}
Let us consider the eigenvalue problem for the matrix $V^L = S^{-1/2}VS^{-1/2}$:
\begin{equation}\label{eigvalU}
  V^L\vec{\upsilon}_k = {u^L_k}\vec{\upsilon}_k
\end{equation}
and construct the matrix of the eigenvectors:
\begin{equation} \label{eigvectmatrU}
  \upsilon = \left( \vec{\upsilon}_1\; \vec{\upsilon}_2\; \vec{\upsilon}_3\;\dots\; \vec{\upsilon}_N \right).
\end{equation}
Since the matrix $V^L$ is Hermitian, the matrix $\upsilon$ is unitary
$(\upsilon^{\dagger} = \upsilon^{-1})$ and the matrix 
\begin{equation} \label{diagonalpotential}
  V^{LV}=\upsilon^{\dagger}V^L\upsilon
\end{equation}
is diagonal. Let us also denote:
\begin{equation} \label{notation3}
  H_0^{LV} = \upsilon^{\dagger}H_0^L{\upsilon}, \;\;
  \vec{C}^{LV} = \upsilon^{\dagger}\vec{C}^L\,.
\end{equation}
With $H_0$ and $S$ substituted from equations (\ref{H0Matrix}) and (\ref{S0Matrix}), respectively, the time-dependent Dirac equation (\ref{NonStat}) takes the form:
\begin{equation} \label{TDDE}
  \dot{\imath}S\dfrac{d}{dt}\vec{C}(t) = \left(H_{0} + \mathcal{F}(t){\cdot}V\right)\vec{C}(t) \,.
\end{equation}
{Multiplying Eq.~(\ref{TDDE}) by $\upsilon^{\dagger}S^{-1/2}$}, we get
\begin{eqnarray} \label{dirac}
 \dot{\imath}\dfrac{d}{dt}\upsilon^{\dagger}S^{1/2}\vec{C}(t)
 &=&\upsilon^{\dagger}S^{-1/2}\left(H_{0} +
\mathcal{F}(t){\cdot}V\right)\nonumber \\
&&\times S^{-1/2}\upsilon\upsilon^{\dagger}S^{1/2}\vec{C}(t)
\end{eqnarray}
or, using the notations (\ref{diagonalpotential}) and (\ref{notation3}),
\begin{equation}
  \dot{\imath}\dfrac{d}{dt}\vec{C}^{LV}(t) = \left(H_{0}^{LV} + \mathcal{F}(t){\cdot}V^{LV}\right)\vec{C}^{LV}(t)\,.
\end{equation}
This equation is suitable for the split-operator method. The short-term propagation can be performed as
\begin{eqnarray}
 \vec{C}^{LV}(t + \Delta{t}) &= & \exp\left[-\dot{\imath}\frac{\Delta{t}}{2}H_0^{LV}\right] \nonumber \\
& \times &\exp\left[-\dot{\imath}\Delta{t}\mathcal{F}\left(t + \dfrac{\Delta{t}}{2}\right)V^{LV}\right]
\nonumber\\
&\times  &\exp\left[-\dot{\imath}\frac{\Delta{t}}{2}H_0^{LV}\right]\vec{C}^{LV}(t) \,,
\end{eqnarray}
where the exponential of $H_0^{LV}$ is obtained by the spectral expansion:
\begin{equation}
  \exp\left[-\dot{\imath}\frac{\Delta{t}}{2}H_0^{LV}\right] = \sum\limits_k\exp\left(-\dot{\imath}
\frac{\Delta{t}}{2}E_k\right)\vec{C}^{LV}_k\left(\vec{C}^{LV}_k\right)^{\dagger}
\end{equation}
and the matrix
\begin{eqnarray}
 \left(\exp\left[-\dot{\imath}\Delta{t}\mathcal{F}\left(t + \dfrac{\Delta{t}}{2}\right)V^{LV}\right]\right)_{ij}
 \nonumber\\
 = \delta_{ij}\exp\left[-\dot{\imath}\Delta{t}\mathcal{F}\left(t + \dfrac{\Delta{t}}{2}\right)u^L_i\right]
\end{eqnarray}
is diagonal (see Eqs. (\ref{eigvalU}) - (\ref{diagonalpotential})).

In order to calculate  the transition and ionization probabilities, we have to project the propagated state onto the vector or the subspace of interest. For instance, to calculate the survival probability in the initial state $\Psi_i$, we have to calculate the scalar product:
\begin{eqnarray}
 \left|\left<\Psi(t)\left|\Psi_i\right.\right>\right|^2 &=& \left|\vec{C}^\dagger(t){\cdot}S\vec{C}_i\right|^2  \nonumber\\ &=&
 \left|\vec{C}^\dagger(t){\cdot}\left(S^{1/2}\right)^{\dagger}\upsilon\upsilon^{\dagger}S^{1/2}\vec{C}_i\right|^2  \nonumber\\ &=& \left|\left(\upsilon^{\dagger}S^{1/2}\vec{C}(t)\right)^\dagger{\cdot}\left(\upsilon^{\dagger}S^{1/2}\vec{C}_i\right)\right|^2 \nonumber\\ &=& \left|\left(\vec{C}^{LV}(t)\right)^\dagger{\cdot}\vec{C}^{LV}_i\right|^2\,.\;
\end{eqnarray}

\begin{figure}[t]
    \includegraphics[width=0.47\textwidth]{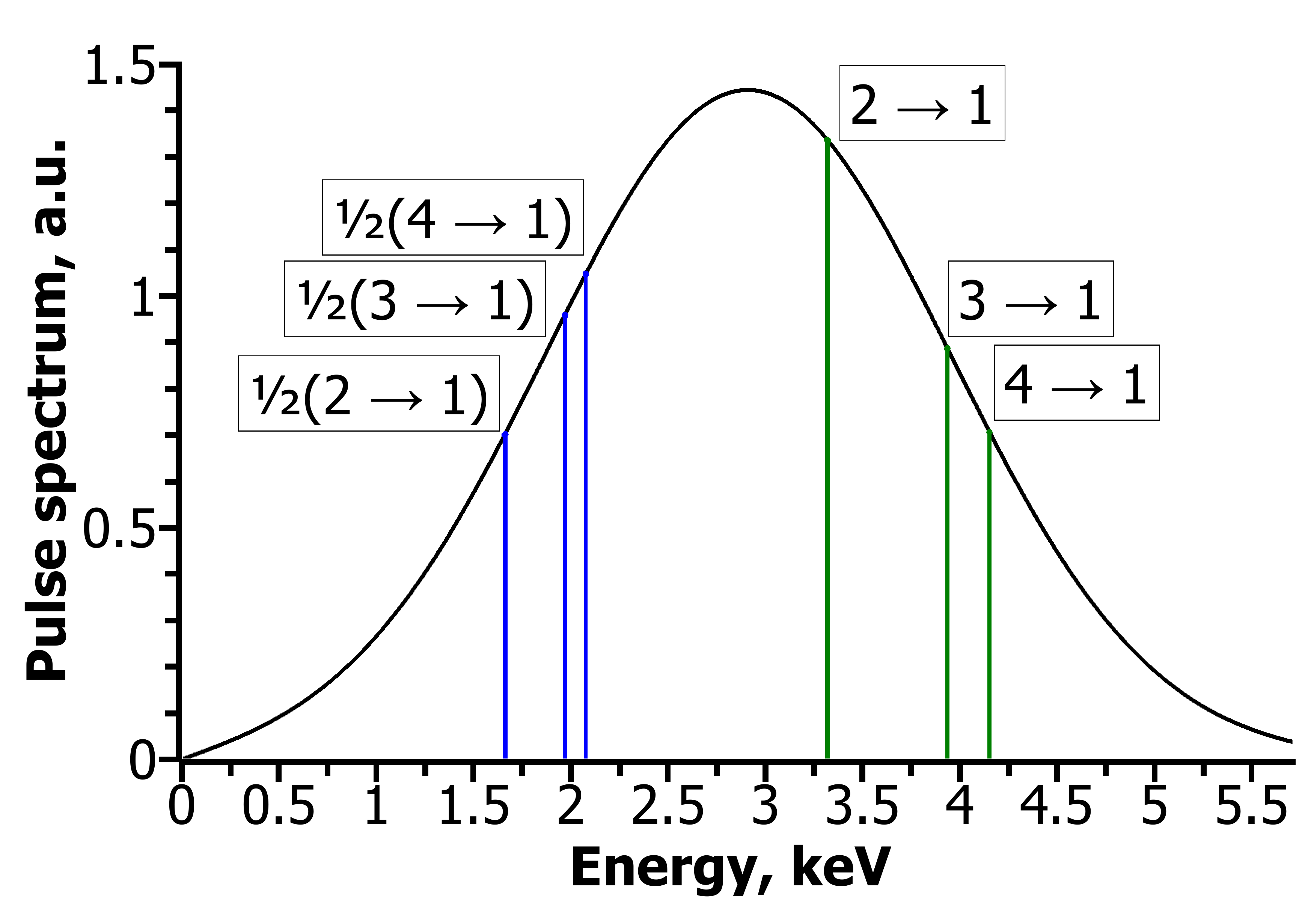}
    \caption{(Color online) The energy spectrum of the Gaussian-shaped laser pulse used in our calculations. Vertical sticks indicate the photon energies necessary for one-photon (green) and two-photon (blue) transitions.
    \label{fouriergausspulse}}
  \end{figure}

\begin{figure}[t]
    \includegraphics[width=0.5\textwidth]{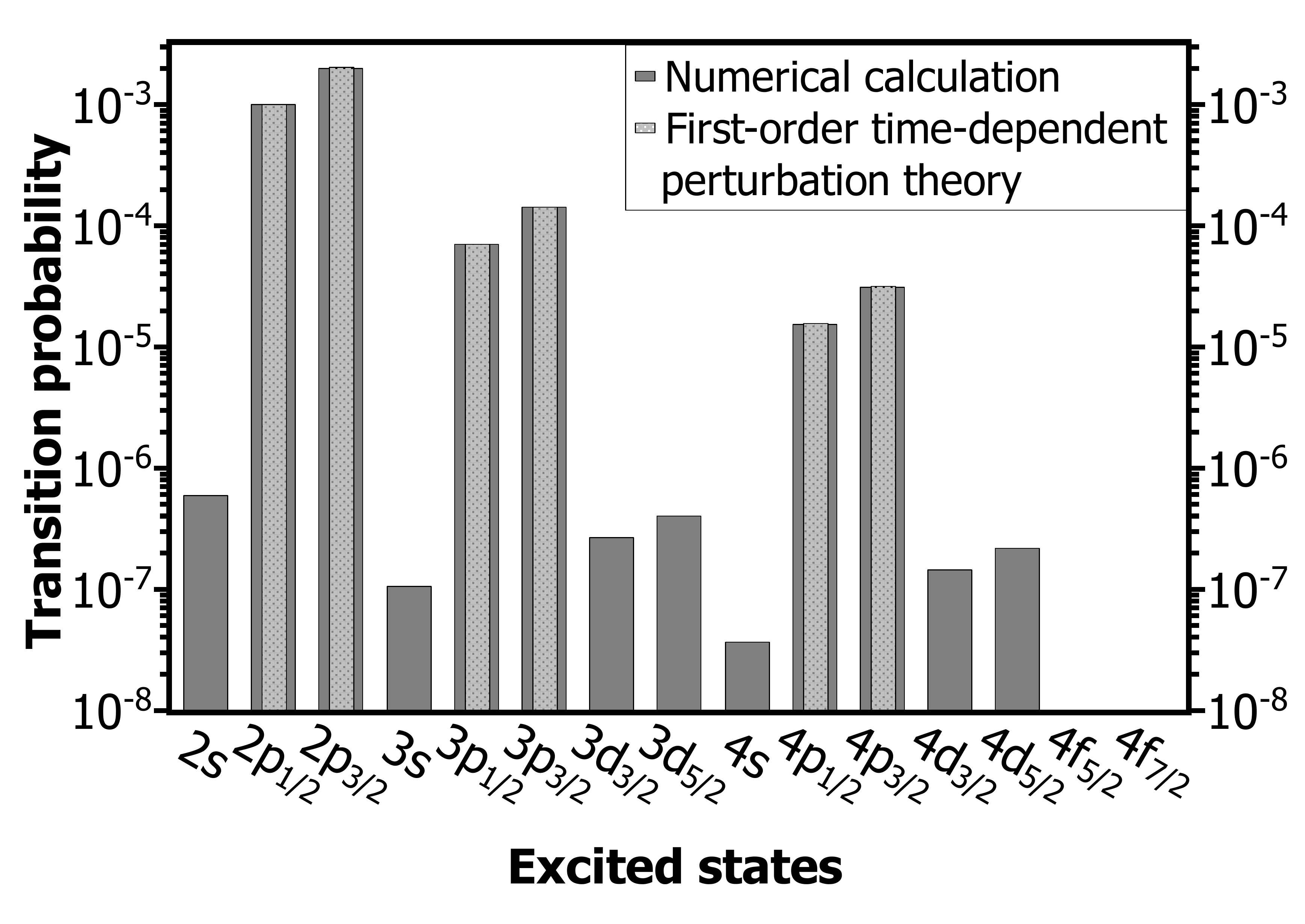}
    \caption{{Electric-dipole transition probabilities from the $1s$ state to the excited states for one-electron argon ion exposed to a Gaussian-shaped laser pulse.} 
    \label{transitions}}
  \end{figure}

The developed methods have been applied to solve two representative problems. First, the transition probabilities in the hydrogenlike argon ion ($Z = 18$) exposed to a short Gaussian-shaped laser pulse are calculated. The form of the pulse is given by the following:
\begin{equation}
  \boldsymbol{\mathcal{F}}(t) = \mathbf{e}_z\mathcal{F}_0 \exp\left(-\dfrac{t^2}{2\tau^2}\right)\sin\left(\omega{t}\right)\,,
\end{equation}
where  $\omega = 4.4$ as$^{-1}$, $\tau = 0.63$ as, and the peak intensity is $I = 6.8{\cdot}10^{19}$ W/cm$^2$.
Fig.~\ref{fouriergausspulse} shows the energy spectrum (i.e. the Fourier transform) of this pulse. We note that the spectrum is broad enough, so that we get not only one-photon transitions, but two-photon ones as well.
In Fig.~\ref{transitions}, we present the transition probabilities from the ground $1s$ state to the excited states due to the interaction with the laser pulse. For comparison, the corresponding results of the first-order time-dependent perturbation theory are shown in the same figure. The initial $1s$ state survival probability is $P_{1s}=0.99666$.
 
Another example is the calculation of the ionization probabilities in the hydrogenlike tin ion ($Z = 50$) exposed to a short $\sin^2$-shaped laser pulse. In this calculation, the pulse is chosen in the form:
\begin{equation}
  \boldsymbol{\mathcal{F}}(t) = \mathbf{e}_z\mathcal{F}_0 \sin^2\left(\dfrac{\pi{t}}{T}\right)\sin\left(\omega{t}\right)\,,
  \;\;\;\;
  t \in [0,T]\,,
\end{equation}
where the wavelength $\lambda$ and the pulse duration $T$ can be expressed through the carrier frequency as $\lambda = 2\pi{c}/\omega$ and $T={2\pi{N}}/{\omega}$. The calculations are performed for $N=~\hspace{-4pt}~20$ and the peak intensity $I=5\times10^{22}$~W/cm$^2$.

\begin{figure}[t]
    \includegraphics[width=0.5\textwidth]{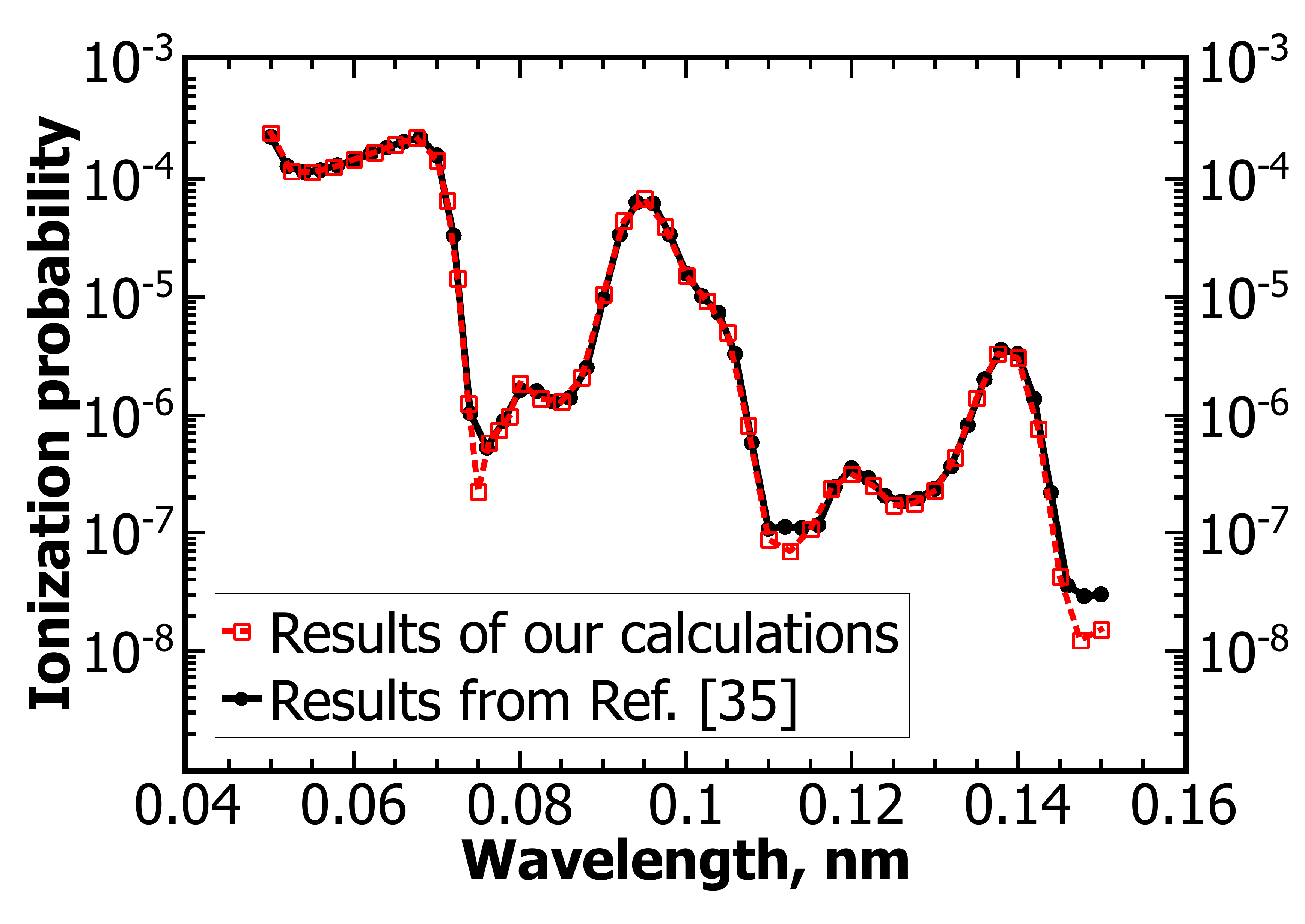}
    \caption{(Color online) Total ionization probability for a tin ion as a function of the laser pulse wavelength. Unshaded red squares connected with the dashed red line: the results of our calculations; shaded black points connected with the solid black line: data taken from Ref. \cite{Saenz}
    \label{graph5}}
  \end{figure}

Fig.~\ref{graph5} displays the full ionization probability as a function of the laser wavelength for all the other parameters of the system kept constant. This plot is in a good agreement with the corresponding data from Ref.~\cite{Saenz}.

\section{Conclusion}
The efficient and easily implementable DKB approach solves the problem of the spurious states related to the use of the finite basis sets for the Dirac equation. In the present paper, this method is generalized for the case of the axial symmetry. Generalized DKB method proved to be accurate and stable in this case. It opens the new way for the fully relativistic theoretical treatment of both stationary and time-dependent axially symmetric problems, e.g., of ions and atoms exposed to external fields. The efficiency of the method is demonstrated by calculating the energies of hydrogenlike ions with non-perturbative account for static uniform external electric or magnetic fields. The Zeeman and Stark energy shifts are compared with the perturbation theory calculations. It is shown that the higher orders of the perturbation theory expansion can be reproduced by the methods developed in the present paper.

For the purpose of solving the time-dependent problem, the finite basis set technique (not regarding the particular choice of the basis set) was adapted to take advantage of the split-operator {method} by the transformation of the {matrix of the} external potential into the diagonal representation. {With this technique, the transition and ionization probabilities for the ions exposed to the laser pulses are evaluated. The results are compared with the corresponding data from other papers or with the independently obtained values. The solution of the time-dependent Dirac equation with the set of the discussed approaches is shown to be correct, accurate and numerically efficient.}

\section{Acknowledgement}
We thank A. S\'{a}enz for providing us with the numerical data from Ref. \cite{Saenz}. Valuable discussions with A. S\'{a}enz and J. L. Sanz-Vicario are gratefully acknowledged. This work was supported in part by~RFBR (Grants No.~12-02-31803 and 13-02-00630), by~the~Ministry of Education and Science of Russian Federation (Project No.~8420), by the FAIR-Russia Research Center, and by the non-profit ``Dynasty'' Foundation.

\end{document}